\documentstyle[12pt,epsfig]{article}
\begin{document}
\setlength{\baselineskip}{0.6cm}
\setlength{\parskip}{0.15cm}

\def \d {{\rm d}}
\def\GE{\gamma_E}
\def\half{\frac{1}{2}}
\def \be {\begin{equation}}
\def \ee {\end{equation}}
\def    \beq             {\begin{equation}}
\def    \eeq             {\end{equation}}
\def    \beqa             {\begin{eqnarray}}
\def    \eeqa             {\end{eqnarray}}
\def \e {{\rm e}}
\def \as {{\alpha_s}}

\begin{flushright}
YITP-SB-00-70\\
BNL-HET-00/42 \\
RBRC-150 \\
November 2000
\end{flushright}

\vspace*{30mm}

\begin{center}
{\LARGE Threshold Resummation and 
Rapidity Dependence}

%\medskip

\par\vspace*{20mm}\par

{\large 
George Sterman$^{a,b}$ and Werner Vogelsang$^c$}

\bigskip

{\em $^a$C.N.\ Yang Institute for Theoretical Physics,
SUNY Stony Brook\\
Stony Brook, New York 11794 -- 3840, U.S.A.}

\bigskip

{\em $^b$Physics Department, Brookhaven National Laboratory,\\
Upton, NY 11973, U.S.A.}

\bigskip

{\em $^c$RIKEN-BNL Research Center, Bldg. 510a, Brookhaven National 
Laboratory, \\
Upton, New York 11973 -- 5000, U.S.A.}

\end{center}
\vspace*{18mm}

\begin{abstract}
We study the effects of threshold resummation on the rapidity dependence
of single-particle-inclusive cross sections, using the prompt photon 
cross section as an example. We make use of the full resummation
formula at next-to-leading logarithmic accuracy and develop a new
technique for treating rapidity in resummation. We compare 
our phenomenological results with those of previous studies and discuss
differences and similarities of the two existing resummation formalisms.
\end{abstract}

\newpage
\section{Introduction}
\noindent
Prompt-photon production at high transverse momentum~\cite{photondata}, 
$pp,p\bar{p},pN\rightarrow \gamma X$, has been a classic tool for 
constraining the nucleon's gluon density, because at leading order a photon 
can be produced in the Compton reaction $qg\to\gamma q$. The ``point-like'' 
coupling of the photon to the quark provides a potentially clean 
electromagnetic probe of QCD hard scattering.
A pattern of disagreement between theoretical predictions and experimental 
data for prompt photon production has been observed in recent 
years~\cite{cdf,e706,ua6}, however,
not globally curable by changing the factorization 
and renormalization scales in the calculation, or by ``fine-tuning'' the gluon 
density~\cite{vv,huston95,aurenche99}. The most serious problems relate
to fixed-target data, where next-to-leading order (NLO) 
theory dramatically underpredicts some data sets~\cite{e706,ua6}. 
At collider energies~\cite{cdf}, there is less reason for concern, 
but also here the agreement is not fully satisfactory. In this context, 
threshold resummations for the inclusive prompt photon cross 
section have been developed in~\cite{LOS,CMN} and applied phenomenologically
in~\cite{CMNOV,KO}. At partonic threshold when the initial partons have 
just enough energy to produce the high-$p_T$ photon and a massless 
recoiling jet, 
the phase space available for gluon bremsstrahlung vanishes, resulting in 
logarithmic corrections to the partonic cross section. Threshold 
resummation~\cite{LOS,CMN,dyresum,BCMN} organizes these corrections 
to all orders in $\as$. 

Phenomenological studies~\cite{CMNOV,KO} of threshold resummation for the 
prompt photon cross section have shown that the size of the resummation 
effects increases with $x_T\equiv 2 p_T/\sqrt{S}$, where $p_T$ is the photon 
transverse momentum. This is expected~\cite{LOS,CMN} from the interplay of the
resummed partonic cross section with steeply falling parton distributions,
and explains why resummation effects are small for the 
prompt photon cross section measured at the Tevatron~\cite{cdf}, where
$x_T$ is generally smaller than 0.1. In contrast to this, for fixed-target 
energies, where values as large as $x_T\approx 0.7$ are attained~\cite{e706}, 
a significant enhancement of the theory prediction was found. 
The enhancements found in~\cite{CMNOV,KO}, however, are 
not sufficient to bring theory in line with all fixed-target
data (or, of course, to reconcile conflicting data sets). 
Nonetheless, valuable insights have been gained from applying threshold 
resummation to the prompt photon cross section. One of them is that 
after resummation the theory predictions show a significant improvement in 
factorization and renormalization scale dependence~\cite{CMNOV} over 
NLO, which has a notoriously large scale sensitivity.
This implies that threshold resummation does provide an important 
addition to the theory result, and has helped develop 
approaches~\cite{LSV,Li} that extend threshold resummation, by 
redistributing its contributions to the hard scattering function 
over soft gluon {\em transverse} momenta, giving rise to further 
enhancement of the cross section through recoil effects on the prompt 
photon spectrum.

We believe that further analyzing the effects of pure 
threshold resummation is useful, for two reasons. First, as 
mentioned above, threshold resummation for the prompt photon cross section 
has been carried out by two groups~\cite{LOS,CMN}. Even though 
both results have been applied numerically~\cite{CMNOV,KO}, a full
comparison of the two formalisms has yet to be carried out, 
either numerically, or at an analytical level. Indeed, it should be
mentioned that there are genuine differences between the two 
formalisms: \cite{LOS} resums at fixed, and arbitrary, photon 
rapidity, while~\cite{CMN} applies to the cross section 
fully integrated over rapidity. Therefore, it is interesting to compare 
the two approaches in cases where they both can be applied, which 
will show the close relation between the two.

Providing phenomenological results for the full content of the resummation 
formula of~\cite{LOS} is our other objective in this paper.
In the numerical study of the inclusive
cross section in~\cite{CMNOV}, the full resummed expression
of~\cite{CMN}, which is at next-to-leading logarithmic (NLL) accuracy, 
was used, although rapidity acceptance was implemented indirectly, 
by adopting the angular dependence of NLO. In contrast, the analysis 
of~\cite{KO}, which made use of the formalism of~\cite{LOS}, 
employed an expansion in the strong coupling
truncated at next-to-next-to-leading order (NNLO). It is therefore
interesting to study the role of higher-order terms, which do 
seem to have a fairly significant influence on the final result.  
To see how, we must develop a new practical technique for treating  
rapidity dependence in the resummed cross section, to which we will 
turn first in Sec.~2. This is an essential extension of the threshold 
resummation formalism, applicable to any resummed cross section at 
measured rapidity. Again, to make direct contact with the numerical 
studies of~\cite{CMNOV}, we adopt the minimal prescription for moment
inversion of~\cite{MP}.

In Sec.~3, we give explicit expressions for the resummation exponents
at NLL accuracy, arising from the formalism of~\cite{LOS}. 
Section~4 presents the NLO hard-scattering functions
that multiply the resummed exponents and collect contributions that are
constant near threshold. The actual comparison between the two resummation
formalisms of~\cite{LOS} and~\cite{CMN} is performed at an analytic
level in Sec.~5. Section~6 discusses our treatment of the inverse
Mellin and Fourier transforms required to obtain the physical cross section
from the resummation formalism laid out in Sec.~2. Finally, in 
Sec.~7 we present numerical results for both resummation formalisms
and provide phenomenological predictions for the prompt photon 
cross section resummed at fixed rapidity.

\section{The resummed prompt-photon cross section}
The cross section $p_T^3 \, \d\sigma/\d p_T \d\eta$ for the single-inclusive 
production of a photon with transverse momentum $p_T$ and 
pseudorapidity $\eta$ is given in terms of the partonic cross section 
$p_T^3 \d\hat{\sigma}/\d p_T \d\eta$ and the distribution functions 
$\phi_{a,b}$ for initial parton types $a,b$ as 
\beq \label{crosec}
\frac{p_T^3 \d\hat{\sigma}}{\d p_T \d\eta} = \sum_{a,b} 
\int_{x_a^{\rm min}}^1 \, dx_a \, \int_{x_b^{\rm min}}^1  \, dx_b \,
\phi_a (x_a,\mu_F^2) \phi_b (x_b,\mu_F^2)  \,
\frac{p_T^3 \d\hat{\sigma}_{ab}}{\d p_T \d\eta}
(x_a P_A,x_b P_B,P_{\gamma},\mu_F,\mu_R) \; ,
\eeq 
where $P_A$, $P_B$, $P_{\gamma}$ are the momenta of the initial
hadrons and the photon, and
\beq
x_a^{\rm min} = \frac{x_T \, {\rm e}^{\eta}}{2-x_T \, {\rm e}^{-\eta}} \;\;\; ,
\;\;\;\;\;\; x_b^{\rm min} = \frac{x_a \, x_T \, {\rm e}^{-\eta}}
{2 x_a-x_T \, {\rm e}^{\eta}} \;\;\; ,
\;\;\;\;\;\; x_T = 2 p_T/\sqrt{s} \; .
\eeq
In Eq.~(\ref{crosec}), $\mu_F$ and $\mu_R$ denote the factorization 
and renormalization scales, respectively.

Threshold resummation is organized in Mellin-$N$ moment
space, where both the evolution of the parton densities and their 
convolution with the hard subprocess cross sections simplify.
Were we interested simply in the cross section $d\sigma/dp_T$, 
integrated over all rapidities, an appropriate variable
for taking moments would be $x_T^2$, as shown in ref.~\cite{CMN}.
If we wish to retain the rapidity dependence of the single-inclusive 
photon cross section $d\sigma/dp_T d\eta$, we need to introduce 
a double transform of the cross section. It turns out to be 
particularly convenient to do this by defining
\begin{equation} \label{doublemom}
\sigma (N,M) \equiv
\frac{1}{\sqrt{2 \pi}} \int_{-\infty}^{\infty} d\eta \, \e^{i M \eta} 
\int_0^{1/\cosh^2 \eta} dx_T^2 \left( x_T^2 \right)^{N-1} 
\frac{p_T^3 \d\sigma}{\d p_T \d\eta} \; .
\end{equation}
In this expression, we take Mellin moments in $x_T^2$ and a Fourier 
transform in rapidity. The latter makes use of the relation between
the hadron and parton level center-of-mass system (c.m.s.) 
rapidities, $\eta$ and $\hat{\eta}$, 
respectively:
\begin{equation}
\eta = \hat{\eta} + \frac{1}{2} \ln \left( \frac{x_a}{x_b} \right)\; ,
\end{equation}
where $x_a,x_b$ are the parton momentum fractions. Then, the Fourier
phase factor $\exp(i M \eta)$ carries through to the parton
level as $\exp(i M \hat{\eta}) \, x_a^{i M/2} x_b^{-i M/2}$, so that we 
will eventually arrive at a Fourier transform in partonic c.m.s.\
rapidity, along with shifts by $\pm iM/2$ in the complex plane of the 
Mellin moments of the parton distributions. Introducing the invariant 
mass of the unobserved hadronic final state, 
\begin{equation}
s_4 = s \left( 1-\hat{x}_T \cosh \hat{\eta} \right) \, ,
\end{equation}
where
\begin{equation} 
s = x_a x_b S \;\; , \;\;\;\; \hat{x}_T =\frac{x_T}{\sqrt{x_a x_b}} \; ,
\end{equation}
we indeed obtain
\begin{eqnarray} \label{smn}
\sigma (N,M)&=&\sqrt{\frac{2}{\pi}} 
\sum_{a,b} \tilde{\phi}_a^{N+1+\frac{i M}{2}}
\tilde{\phi}_b^{N+1-\frac{i M}{2}} \int_{-\infty}^{\infty} d\hat{\eta} \,
\e^{i M \hat{\eta}} \left( \cosh \hat{\eta} \right)^{-2 N} 
\int_0^{s} \frac{\d s_4}{s} 
\left(1-\frac{s_4}{s} \right)^{2N-1} 
\frac{p_T^3 \d\hat{\sigma}_{ab}}{\d p_T \d\hat{\eta}} \nonumber \\
&\equiv& \frac{1}{\sqrt{2\pi}} \sum_{a,b} \tilde{\phi}_a^{N+1+\frac{i M}{2}}
\tilde{\phi}_b^{N+1-\frac{i M}{2}} \int_{-\infty}^{\infty} d\hat{\eta} \,
\e^{i M \hat{\eta}} \left( \cosh \hat{\eta} \right)^{-2 N} 
\tilde{\omega}_{ab} (2N,\hat{\eta}) \; .
\end{eqnarray}
The second relation serves as the definition of the hard-scattering
function $\tilde{\omega}_{ab}$, in Mellin-moment space, as a function 
of rapidity. Moments of the parton distributions are denoted by 
\begin{equation}
\tilde{\phi}_a^L \equiv \int_0^1 dx \; x^{L-1} \; \phi_a(x,\mu_F^2) \; .
\end{equation}

Threshold resummation for the single-particle inclusive cross section 
organizes corrections as singular as $\alpha_s^n \left[ \ln^m 
\left( s_4/s \right)/s_4 \right]_+$, $m\leq 2n-1$,
near partonic threshold, $s_4\rightarrow 0$,
to all orders in perturbation theory. Let us define by $N'\equiv 2 N$
the moment variable conjugate to $(1-s_4/s)$. This variable was
denoted as $N$ in Ref.~\cite{LOS}. In Mellin-$N'$ space,
these logarithms translate into terms of the form $\alpha_s^n 
\ln^m (N')$, $m\leq 2n$. The hard scattering functions 
$\tilde{\omega}_{ab} (N',\hat{\eta})$ for $qg$ and $q\bar{q}$ scattering, 
resummed to NLL accuracy at {\em fixed} 
rapidity $\hat{\eta}$, were derived in~\cite{LOS}, and are given as
\beqa
\tilde{\omega}_{ab} (N',\hat{\eta}) 
&=& \alpha \, \alpha_s(\mu_R^2) \,\; 
C_{ab} \left( \hat{\eta},\alpha_s(\mu_R^2)\right) \, \;
\exp \left \{E'_c (N',s) \right\}
\exp \left \{ \sum_{i=a,b} E_i (N_i',s)
\right \} \nonumber\\
&\ & \hspace{3mm} \times
\exp \left[\int_{\mu_R}^{\sqrt{s}/N'} {d\mu' \over \mu'} \, 
2 \, {\rm Re} \Gamma_S^{(ab\to  \gamma c)}
\left(\hat \eta,\alpha_s(\mu'^2)\right)\right]\, ,
\label{rescrosec}
\eeqa 
with $N_i' = 2 N_i$, $N_a=N (-u/s)$, $N_b=N (-t/s)$.
We now describe each of these functions, starting from the left.
The coefficients $C_{ab} (\hat{\eta},\alpha_s)$ include the 
underlying Born $2\to 2$ hard scatterings $ab \to \gamma c$, and also 
absorb factors in the cross section that are constant in the 
large-$N'$ limit at fixed $\hat{\eta}$, 
produced by hard virtual and non-logarithmic soft 
higher-order contributions~\cite{CMN}. The $C_{ab}$ are subject to
$2\to 2$ kinematics, with $\hat{x}_T = 1/\cosh (\hat{\eta})$, and
are functions of $\hat{\eta}$ and the strong coupling only,
\beq
C_{ab} (\hat{\eta},\alpha_s) = \sum_{n=0}^{\infty}
\left( \frac{\alpha_s}{\pi} \right)^n C_{ab}^{(n)} (\hat{\eta}) \; .
\label{cexp}
\eeq
Their leading terms $C_{ab}^{(0)}$ are just the $2\to 2$ cross sections 
for $ab \to \gamma c$, given in our normalization by
\beqa \label{clo}
C_{q\bar{q}}^{(0)} \left(\hat{\eta}\right) &=& 
\frac{\pi e_q^2 C_F}{C_A} \frac{1}{2 \cosh^4 (\hat{\eta})}
\left( \frac{t}{u}+\frac{u}{t} \right) \; , \nonumber \\
C_{qg}^{(0)} \left(\hat{\eta}\right) &=& 
\frac{\pi e_q^2 }{2C_A} \frac{1}{2 \cosh^4 (\hat{\eta})}
\left( - \frac{t}{s}-\frac{s}{t} \right) \; , 
\eeqa
where $t=(p_a-p_{\gamma})^2 = -s \, \hat{x}_T \, {\rm e}^{-\hat{\eta}}/2$, 
$u=(p_b-p_{\gamma})^2 = - s \, \hat{x}_T \, {\rm e}^{\hat{\eta}}/2$.
The next-order terms $C_{ab}^{(1)}$ can be extracted explicitly~\cite{CMN} 
from the complete NLO calculation of~\cite{gv}; see below. 

To the right of the coefficient functions $C_{ab}$
in Eq.~(\ref{rescrosec}), each of the exponential factors is 
associated with one of the sets of on-shell quanta that characterize
the hard-scattering process near partonic threshold. These are:
quanta collinear to the incoming partons, quanta collinear 
to the final-state jet that recoils against the photon, and soft-gluon
radiation.

The resummed exponents for the initial state partons in Eq.~(\ref{rescrosec})
are given by
\beqa
E_i (2N_i,M_i) &=& -\int^1_0 dz \frac{z^{2N_i-1}-1}{1-z}\;
\left \{\int^1_{(1-z)^2} \frac{dt}{t}
A_i \left[\alpha_s\left(t M_i^2 \right)\right]
+\bar B_i\left(\nu_i,{M_i^2\over s},\alpha_s((1-z)^2 M_i^2 )\right)
 \right \} 
\nonumber\\
&\ & \hspace{3mm}
-2\int_{\mu_R}^{M_i}{d\mu'\over\mu'}\; 
\gamma_i\left( \alpha_s(\mu'{}^2) \right)+
2\int_{\mu_F}^{M_i}{d\mu'\over\mu'}\; \gamma_{ii}
\left( 2N_i,\alpha_s(\mu'{}^2)\right) \; ,
\label{Eexp}
\eeqa
where $M_i$ is a scale of order $\sqrt{s}$. As we shall
see, the exponent is actually independent of $M_i$ to NLL.
In addition,
\beqa
&&A_i (\alpha_s) = C_i \left ( {\alpha_s\over \pi}
+\frac{1}{2} K \left({\alpha_s\over \pi}\right)^2\right )\, , \;\;\;\;
K= C_A\; \left ( {67\over 18}-{\pi^2\over 6 }\right ) - {5\over 9}n_f \; ,
\nonumber \\
&&\bar B_i\left(\nu_i,{M_i^2\over s},\alpha_s\right)
=C_i\; {\alpha_s\over\pi} \left[1-\ln (2 \nu_i) +\ln 
\left( \frac{M_i^2}{s} \right) \right] \, .
\label{anuf}
\eeqa
Here, $C_q = C_F=4/3$ and $C_g = C_A=3$, and $n_f$ is the number of flavors.
The $\nu_i$ contain the gauge dependence, introduced to define the
factorized cross section~\cite{LOS},
\beq
\nu_i \equiv \frac{(\beta_i \cdot n)^2}{|n^2|} \, .
\label{nudef}
\eeq
where $\beta_i^{\mu}=
p_i^{\mu} {\sqrt {2/s}}$, and $n$ an axial gauge vector. Furthermore,
to one loop order 
\beqa
\gamma_q (\as)&=& {3\over 4} C_F {\alpha_s\over\pi}\, , \qquad
\gamma_{qq} (N,\as)= - \left(\ln{N} - {3\over 4}\right) 
C_F {\alpha_s\over\pi} \, ,
\nonumber \\ \gamma_g (\as)&=& b_0 \alpha_s\, , \qquad
\gamma_{gg} (N,\as)= - \left(C_A \ln{N}-\pi b_0\right){\alpha_s\over\pi}\, ,
\eeqa
where $b_0= (11 C_A-4 T_R n_f)/(12\pi)$, with $T_R=1/2$,
and where $\bar N\equiv N{\rm e}^{\gamma_E}$,
with $\gamma_E$ the Euler constant. The $\gamma_i$ are the
anomalous dimensions of the quark and gluon fields, and the 
$\gamma_{ii}$ the $\ln N$ and constant terms in the negative of the 
moments of the diagonal splitting functions. 
As can be seen from (\ref{rescrosec}), 
they organize renormalization- and factorization-scale dependence,
respectively.  

The four terms in the initial-state exponent
are closely linked by the relation between the large-$N$
behavior of diagonal splitting functions and anomalous dimensions,
\beqa
\gamma_{ii}(N,\alpha_s) = - \ln \bar N\, A_i(\alpha_s) + \gamma_i(\as)\, ,
\label{iiAi}
\eeqa
where, as above, $\bar N= Ne^{\gamma_E}$.
Each of the four terms in $E_i$
plays a specific role in the resummation process.  
The first, double-integral, term organizes double-logarithmic behavior
associated with soft and collinear divergences
in terms of the universal behavior of splitting functions. The second,
single-logarithmic, term matches the behavior
of the splitting functions to that of the physical cross section,
factorized as in Ref.~\cite{LOS}. The third term summarizes  noneikonal
logarithms in virtual diagrams, while the fourth matches
large transverse
momentum subtractions associated with the $\overline{\rm MS}$
scheme to the double-logarithmic term.  
Using Eq.\ (\ref{iiAi}), we readily show that at NLL, $E_i$ is
independent of the $M_i$. In our notation, Refs.~\cite{LOS} 
and~\cite{KO} adopt the choices $M_a = -u/\sqrt{s}$ and 
$M_b = -t/\sqrt{s}$.

From the final-state jet that recoils against the photon, one has the exponent
\beqa
E'_c (2N,s) &=& \int^1_0 dz \frac{z^{2N-1}-1}{1-z}\;
\Bigg \{\int^{(1-z)}_{(1-z)^2} \frac{dt}{t}
A_c \left[\alpha_s(t s)\right] 
-\gamma_c\left[\alpha_s((1-z)s) \right] 
\nonumber\\
&\ & \hspace{5mm} 
-\bar B_c\left[\nu_c,1,\alpha_s((1-z)^2s) \right]\Bigg \}
+2  \int_{\mu_R}^{\sqrt{s}} {d\mu'\over \mu'}
\gamma_c(\alpha_s(\mu'{}^2))\;
\, ,
\label{Eprexp}
\eeqa
where $\bar B_c$ is defined in Eq.\ (\ref{anuf}) above.
As for the initial-state jet, the $A_c$ and $\bar B_c$
terms may be associated with the eikonal approximation,
while the $\gamma_c$ terms organize the remaining,
collinear logarithms.  Again, compared to Ref.\ \cite{LOS},
we make the dependence on the renormalization scale
explicit, although it is not associated with logarithms
of the moment variable.

The final factor in Eq.\ (\ref{rescrosec}) resums 
NLL terms from coherent soft gluon emission
between the jets.  The rapidity-dependent
anomalous dimension governing this soft radiation 
depends on the flavor combination in the hard scattering.
For
the process $q{\bar q}\rightarrow  \gamma g$, it is given by
\beqa \label{gamqq}
\Gamma_S^{(q {\bar q} \rightarrow \gamma g)}(\hat \eta,\as)
=\frac{\alpha_s}{2\pi}
\left\{C_F\left[-\ln(4\nu_q \nu_{\bar q})+2 -2 \pi i\right] 
+C_A\left[\ln \left(\frac{tu}{s^2}\right)
+1 -\ln(2 \nu_g) +\pi i \right]\right\} \, .
\eeqa
For the process $qg \rightarrow \gamma q$, it reads
\beqa
\Gamma_S^{(qg \rightarrow \gamma q)}(\hat \eta,\as)=\frac{\alpha_s}{2\pi}
\left\{C_F\left[2\ln \left(\frac{-u}{s}\right)
-\ln(4\nu_{q_a} \nu_{q_c})+2 \right] 
+C_A\left[\ln\left(\frac{t}{u}\right)
-\ln(2 \nu_g) +1 -\pi i \right]\right\} \, .
\eeqa
Note that the dependence on the gauge vector $n$ in all these expressions, 
contained in the $\nu_i$, Eq.~(\ref{nudef}), 
drops out when all contributions to 
Eq.~(\ref{rescrosec}) are combined~\cite{LOS,KO,KOS}.
Indeed, by redefining the soft and jet functions
as in Refs. \cite{LSV}, we may eliminate
this dependence at the beginning.  Here, however,
we retain it, again for ease of comparison to Refs. \cite{LOS}
and \cite{KO}.

\section{NLL exponents}
Following the procedures of~\cite{CMN,CT}, we reexpress the three exponents in
Eq.~(\ref{rescrosec}) in a form with NLL accuracy. The respective 
results for each exponent are given in Appendix~A. The sources of 
$\hat{\eta}$ dependence are readily visible from the expressions given 
there and are: the dependence of the initial-state exponents $E_i$ on
$\lambda_i\equiv \as b_0 \ln N_i$ (where $\alpha_s \equiv \alpha_s (\mu_R^2)$),
the dependence of the final-state jet function on $s=4 p_T^2 \cosh^2 \hat
\eta$, and the dependence of the soft anomalous dimensions 
$\Gamma_S^{(ab\to\gamma c)}$ on the subprocess  Mandelstam variables
$s,t,u$, as given above. After 
combining all NLL exponents given in Appendix~B, 
we find that for both partonic channels the $\hat{\eta}$ dependence 
is purely of the form $\ln (\cosh \hat{\eta})$:
\beqa \label{etaexp}
&&
\hspace*{-0.9cm}
\left( \alpha \, \alpha_s \right)^{-1} \; \int_{-\infty}^{\infty} d\hat{\eta} \,
\e^{i M \hat{\eta}} \left( \cosh \hat{\eta} \right)^{-2 N} \tilde{\omega}_{ab} 
(2N,\hat{\eta}) \\
&\hspace*{-0.3cm}=&\hspace*{-0.3cm}
\int_{-\infty}^{\infty} d\hat{\eta} \,
\e^{i M \hat{\eta}} \left( \cosh \hat{\eta} \right)^{-2 N}
C_{ab} (\hat{\eta},\as) 
\exp \left[ \ln \tilde{N} g_{ab}^{(1)} (\tilde{\lambda}) + 
g_{ab}^{(2)} \left(\tilde{\lambda},\frac{Q^2}{\mu_R^2},
\frac{Q^2}{\mu_F^2}\right) +
\ln (\cosh \hat{\eta}) \, g_{ab}^{(3)} (\tilde{\lambda}) \right] \nonumber \\
&\hspace*{-0.3cm}=&\hspace*{-0.3cm} 
\exp \left[ \ln \tilde{N} 
g_{ab}^{(1)} (\tilde{\lambda}) + g_{ab}^{(2)} \left(\tilde{\lambda},
\frac{Q^2}{\mu_R^2},\frac{Q^2}{\mu_F^2}\right)\right]
\int_{-\infty}^{\infty} d\hat{\eta} \, \e^{i M \hat{\eta}} 
\left( \cosh \hat{\eta} \right)^{-2 N + g_{ab}^{(3)} (\tilde{\lambda})}
\left[ C_{ab}^{(0)} (\hat{\eta}) + \frac{\alpha_s}{\pi} 
 C_{ab}^{(1)} (\hat{\eta}) \right] \; ,\nonumber
\eeqa
where $Q^2\equiv 2 p_T^2$, and $\tilde{N}\equiv N+1$,
$\tilde{\lambda} \equiv  \alpha_s \, b_0 \ln (N+1)$. Here we have adopted
the convention of~\cite{CMN} to evaluate the resummation exponents at 
the moment variable $N+1$, rather than at $N$, since the former naturally  
appears in the moments of the parton
densities; see Eq.~(\ref{smn}) and Eqs.~(\ref{comp}), (\ref{cmncomp})
below. The functions $g_{ab}^{(1)} (\lambda)$ and $g_{ab}^{(2)} (\lambda,
Q^2/\mu_R^2,Q^2/\mu_F^2)$ (where $\lambda \equiv  \alpha_s \, b_0 \ln N$) 
resum LL and NLL terms, respectively, that are independent of 
$\hat{\eta}$. Resummation of the $\hat{\eta}$ dependence is taken care of by 
the functions $g_{ab}^{(3)} (\lambda)$, which are of NLL order only. 
The particular simplicity of the $\hat{\eta}$ dependence in the exponent 
makes it possible to join all $g_{ab}^{(3)}$
terms with the moment variable in the $\hat{\eta}$ integrand. 

The functions $g_{ab}^{(1)} (\lambda)$ and $g_{ab}^{(2)} (\lambda,
Q^2/\mu_R^2,Q^2/\mu_F^2)$
have already been presented in~\cite{CMN}, and we recall them here: 
\beqa
g_{q{\bar q}}^{(1)}(\lambda) &=& (2C_F - C_A) \;h^{(1)}(\lambda) +
C_A \;h^{(1)}(\lambda/2) \;, \;\;\; \;\;
\nonumber \\\label{g1fun}
g_{qg}^{(1)}(\lambda) &=& C_A \;h^{(1)}(\lambda) +
C_F \;h^{(1)}(\lambda/2) \;, 
\eeqa
where
\beq
\label{htll}
h^{(1)}(\lambda) =
\frac{1}{2\pi b_0 \lambda}\Bigl[ 2\lambda + (1-2\lambda)
\ln(1-2\lambda) \Bigr] \; ,
\eeq
while, 
\beqa
\label{g2qq}
g_{q{\bar q}}^{(2)}\!\left(\lambda,\frac{Q^2}{\mu_R^2},
\frac{Q^2}{\mu_F^2}\right)
&=& (2C_F - C_A) \;h^{(2)}(\lambda) + 2 \,C_A \;h^{(2)}(\lambda/2)  \\
&+& \frac{2C_F - C_A}{2\pi b_0} \ln 2 \ln(1-2\lambda) 
+ \frac{ C_A \GE -  \pi b_0}{\pi b_0} \ln(1-\lambda) 
- \frac{2C_F}{\pi b_0} \;\lambda \ln \frac{Q^2}{\mu_F^2}
\nonumber \\
&+& \left\{ \frac{C_F}{\pi b_0} \Bigl[ 2\lambda + \ln(1-2\lambda) \Bigr]  
+ \frac{C_A}{2\pi b_0} \Bigl[ 2 \ln(1-\lambda) - \ln(1-2\lambda) \Bigr]
\right\} \ln \frac{Q^2}{\mu_R^2} \;, \nonumber \\
\label{g2qg}
g_{qg}^{(2)}\!\left(\lambda,\frac{Q^2}{\mu_R^2},\frac{Q^2}{\mu_F^2}\right)
&=& C_A \;h^{(2)}(\lambda) + 2 \,C_F \;h^{(2)}(\lambda/2) \\
&+& \frac{C_A}{2\pi b_0} \ln 2 \ln(1-2\lambda) 
+ \frac{4 C_F \,\GE - 3 C_F}{4\pi b_0} \ln(1-\lambda) 
- \frac{C_F + C_A}{\pi b_0} \;\lambda \ln \frac{Q^2}{\mu_F^2}
\nonumber \\
&+& \left\{ \frac{C_F + C_A}{2\pi b_0} \Bigl[ 2\lambda + 
\ln(1-2\lambda) \Bigr]  
+ \frac{C_F}{2\pi b_0} \Bigl[ 2 \ln(1-\lambda) - \ln(1-2\lambda) \Bigr]
\right\} \ln \frac{Q^2}{\mu_R^2} \;, \nonumber
\eeqa
where $Q^2=2 p_T^2$, $\GE$ is the Euler constant, $\GE = 0.5772\ldots$, and
\beq
\label{htnll}
h^{(2)}(\lambda) =
\frac{b_1}{2\pi b_0^3}\Bigl[ 2\lambda + \ln(1-2\lambda)
+\half\ln^2(1-2\lambda) \Bigr] 
- \frac{\GE}{\pi b_0} \ln(1-2\lambda) 
- \frac{K}{4\pi^2 b_0^2}\Bigl[ 2\lambda + \ln(1-2\lambda)\Bigr] 
\; , 
\eeq
with $b_1 = (17C_A^2 -10 \, C_A T_R n_f-6 \, C_F T_R n_f)/(24\pi^2)$. 
The coefficient $K$ has already been defined in~(\ref{anuf}).
The functions $g_{ab}^{(3)} (\lambda)$ are new to this analysis and read
\begin{eqnarray}
g_{q{\bar q}}^{(3)}(\lambda) &=& (2C_F - C_A) \;h^{(3)}(\lambda) +
C_A \;h^{(3)}(\lambda/2) \;, \;\;\; \;\;
\nonumber \\\label{g3fun}
g_{qg}^{(3)}(\lambda) &=& C_A \;h^{(3)}(\lambda) +
C_F \;h^{(3)}(\lambda/2) \;, 
\end{eqnarray}
where 
\begin{equation} \label{g3fun1}
h^{(3)}(\lambda) = \frac{2}{\pi b_0} \ln(1-2\lambda) \; . 
\end{equation}
As mentioned earlier, they only contribute at NLL level. 

Finally, following Eq.~(\ref{cexp}), we have expanded the hard-scattering 
functions $C_{ab} (\hat{\eta},\as)$ in 
Eq.~(\ref{etaexp}) to first order in $\alpha_s$. As was pointed out
in~\cite{CMN,KO}, the inclusion of the $C_{ab}^{(1)}$ terms is significant,
since with their help {\em three} towers of logarithms, $\alpha_s^n 
\ln^m (N)$, $2n-2\leq m\leq 2n$, are brought under control in 
Eqs.~(\ref{rescrosec}), (\ref{etaexp}). In this way, the factors $C_{ab}$ 
also help reduce the dependence of the cross section on the renormalization 
scale $\mu_R$. 

\section{The hard-scattering functions $C_{ab}^{(1)}$}
Expanding the expressions above in $\alpha_s$, and comparing them to the
exact NLO calculation of~\cite{gv}, we are able to check that 
indeed the logarithms of the type $\alpha_s \ln^2 (N)$ and 
$\alpha_s \ln (N)$, which appear at NLO, are correctly reproduced 
by the resummation calculation, including their rapidity dependence~\cite{KO}. 
This comparison
also determines the explicit expressions for the $N$-independent 
coefficients $C_{ab}^{(1)} (\hat{\eta})$. For the 
$q\bar{q}$ annihilation process, we find:
\begin{eqnarray} \label{c1qqb}
C_{q\bar{q}}^{(1)} (\hat{\eta}) &=& 
\frac{1}{96} \, 
C_{q\bar{q}}^{(0)} (\hat{\eta}) \left \{ 
12 C_F \left( 2 \ln [v (1-v)]  \Big( 1+\ln [v (1-v)] \Big) -
4 \rho_{q\bar{q}} \ln \left[ \frac{Q^2}{2 \mu_F^2 v (1-v)} \right] 
\right.\right.
\nonumber \\
&&\hspace*{4.4cm} -4 \ln v \ln (1-v) +\frac{20}{3} \pi^2 +
\rho_{q\bar{q}}^2+6 \rho_{q\bar{q}}-19 \Bigg) \nonumber \\
&&\hspace*{1.5cm}-C_A  \Bigg(12 (1+\rho_{q\bar{q}}) \ln[v (1-v)] 
+24 \Big( \ln v \ln (1-v) +\pi^2 \Big) +
3 \rho_{q\bar{q}}^2+18 \rho_{q\bar{q}}-5 \Bigg) \nonumber \\
&&\left. \hspace*{1.5cm}+8 \pi b_0 \left( 29+3 \rho_{q\bar{q}} -12 
\ln \left[ \frac{Q^2}{2 \mu_R^2 \sqrt{v (1-v)}} \right] \right)
\right\} 
\nonumber \\
&& + \frac{\pi e_q^2 C_F}{C_A}  v (1-v) \Bigg\{ - (C_A-2 C_F) 
 (1+2 v)\ln^2 v - 2 C_A \ln v + 2 C_F  \Big(5 - 6 v \Big) \ln v \nonumber \\
&& \hspace*{3.4cm}
+ \Big(v \longleftrightarrow 1-v \Big)\Bigg\} \; ,
\end{eqnarray}
where 
\beqa \label{rhqqb}
v &=& 1 + \frac{t}{s} = \frac{{\rm e}^{\hat{\eta}}}
{2\cosh \hat{\eta}} \nonumber \\
\rho_{q\bar{q}} &=& -3+4 \gamma_E + 2 \ln [4 v (1-v)] \; .
\eeqa
For $qg$ QCD Compton scattering, one has:
\begin{eqnarray}\label{c1qg}
C_{qg}^{(1)} (\hat{\eta}) &=& 
\frac{1}{4} \, C_{qg}^{(0)} (\hat{\eta}) \left \{ 
C_F \left( 4 \ln^2 \frac{1-v}{v} +2 (\ln v-3) \ln v 
-\rho_{qg}^{(F)} \ln \left[ \frac{Q^2}{2 \mu_F^2 (1-v)} 
\right] \right. \right.  
\nonumber \\
&&\hspace*{4.4cm} +\frac{1}{8}\left( \rho_{qg}^{(F)} \right)^2
+\frac{3}{2} \rho_{qg}^{(F)}
+\frac{11}{3}\pi^2 -\frac{29}{8} \Bigg) \nonumber \\
&&\hspace*{1.5cm}-\left. C_A \left( \ln^2 \frac{1-v}{v} +
\rho_{qg}^{(A)} \ln \left[ \frac{Q^2}{2 \mu_F^2 v^2} \right]
-\frac{1}{4} \left( \rho_{qg}^{(A)} \right)^2
+\frac{1}{3} \pi^2 \right)-4 \pi b_0 \ln \left[ \frac{\mu_F^2}{\mu_R^2}
\right] \right\} \nonumber \\
&&+\frac{\pi e_q^2}{C_A} v^2 (1-v) \Bigg\{
(C_A-2 C_F) \left[ (3-2 v) \left( \ln^2 \frac{1-v}{v} + \pi^2 \right) 
+ (1-2 v) \ln^2 v \right.\nonumber \\
&&\hspace*{3.4cm}\left. + 2 (1-v)  \ln \frac{1-v}{v^2}  \right] +
6 C_F \ln (1-v) \Bigg\} \; ,
\end{eqnarray}
where
\beqa \label{rhoa}
\rho_{qg}^{(F)} &=& -3 +4 \Big( \gamma_E + \ln [2v] \Big) \nonumber \\
\rho_{qg}^{(A)} &=& 4 \Big( \gamma_E + \ln [2(1-v)] \Big) \; .
\eeqa
The values of the ratios $C_{ab}^{(1)}(\hat{\eta})/C_{ab}^{(0)}(\hat{\eta})$ at
$\hat{\eta}=0$ (that is, at $v=1/2$) agree with the corresponding 
expressions in~\cite{CMN}. Despite the somewhat 
complicated rapidity dependence of the $C_{ab}^{(1)}$, it is possible to
evaluate the integral in the last line of Eq.~(\ref{etaexp}) 
in closed form. These results are given in Appendix~B and 
complete the derivation of the resummed photon cross 
section in Mellin-$N$ and Fourier space. We note that the explicit 
$\GE$-terms of the coefficients, in Eqs.~(\ref{rhqqb}) and (\ref{rhoa}),
as well as in $g^{(2)}_{ab}$ in Eqs.~(\ref{g2qq}) and (\ref{g2qg}),
can be eliminated~\cite{Nick00} by redefining $\lambda =\as b_0 \left(\ln N  
{\rm e}^{\GE} \right)\equiv \as b_0 \ln \bar{N} $. 
This convention was adopted in Ref.~\cite{LSV}.

\section{The rapidity-integrated resummed cross section}
We now compare to the 
calculation presented in~\cite{CMN}, which aimed at 
threshold resummation of the $\eta$-{\em integrated} cross section. 
According to Eq.~(\ref{doublemom}), integration over $\eta$ is equivalent
to setting $M=0$ in our expressions and multiplying
by $\sqrt{2\pi}$. We then find 
from~(\ref{doublemom}), (\ref{smn}), (\ref{etaexp})
for the resummed cross section:
\beqa \label{comp}
&&\hspace*{-0.7cm} \int_{-\infty}^{\infty} d\eta \, 
\int_0^{1/\cosh^2 \eta} dx_T^2 \left( x_T^2 \right)^{N-1} 
\frac{p_T^3 \d\sigma}{\d p_T \d\eta}  \\
&\hspace*{-0.4cm}=&\hspace*{-0.2cm} 
\alpha \, \alpha_s \, \sum_{a,b} \tilde{\phi}_a^{N+1} \tilde{\phi}_b^{N+1} 
\exp \left[ \ln \tilde{N} 
g_{ab}^{(1)} (\tilde{\lambda}) + g_{ab}^{(2)} (\tilde{\lambda}) \right]
\int_{-\infty}^{\infty} d\hat{\eta} \,
\left( \cosh \hat{\eta} \right)^{-2 N + g_{ab}^{(3)} (\tilde{\lambda})}
\left[ C_{ab}^{(0)} (\hat{\eta}) + \frac{\alpha_s}{\pi} 
C_{ab}^{(1)} (\hat{\eta}) \right]
\; ,\nonumber
\eeqa
to be compared to the result of~\cite{CMN}, which can be written 
as
\beqa
\label{cmncomp}
&&\hspace*{-0.9cm} \int_{-\infty}^{\infty} d\eta \, 
\int_0^{1/\cosh^2 \eta} dx_T^2 \left( x_T^2 \right)^{N-1} 
\frac{p_T^3 \d\sigma}{\d p_T \d\eta}  \\
&\hspace*{-0.3cm}=&\hspace*{-0.2cm} 
\alpha \, \alpha_s \, \sum_{a,b} \tilde{\phi}_a^{N+1} \tilde{\phi}_b^{N+1} 
\exp \left[ \ln \tilde{N} 
g_{ab}^{(1)} (\tilde{\lambda}) + g_{ab}^{(2)} (\tilde{\lambda}) \right]
\left[ 1 + \frac{\alpha_s}{\pi} 
\frac{C_{ab}^{(1)} (0)}{C_{ab}^{(0)} (0)} \right]
\int_{-\infty}^{\infty} d\hat{\eta} \,
\left( \cosh \hat{\eta} \right)^{-2 N} \, C_{ab}^{(0)} (\hat{\eta})
\; . \nonumber
\eeqa
The two results become identical if the resummed exponent and 
the first-order correction in $C_{ab}$ in Eq.~(\ref{comp}) are evaluated at 
$\hat{\eta}=0$, instead of retaining their full $\hat{\eta}$ 
dependence. We note from~(\ref{comp}), however, 
that for large $N$ the integrand 
is indeed strongly dominated by the region $\hat{\eta} \approx 0$. 
Therefore, setting $C_{ab}^{(1)}(\hat{\eta})/C_{ab}^{(0)}(\hat{\eta})
= C_{ab}^{(1)}(0)/C_{ab}^{(0)}(0)$
and neglecting the term proportional to $g_{ab}^{(3)}$
is expected to be a good approximation, and
we anticipate that the two resummation formalisms should yield
numerically very similar results for the $\eta$-integrated resummed 
cross section. On the other hand, if one wants to obtain fully 
consistent resummed results at fixed rapidity, one needs 
to make use of the full expression in Eq.~(\ref{etaexp}) resulting 
from the formalism of~\cite{CMN}. 

\section{Inverse transforms}
\begin{figure}[h]
\hspace*{-2cm}
\epsfig{file=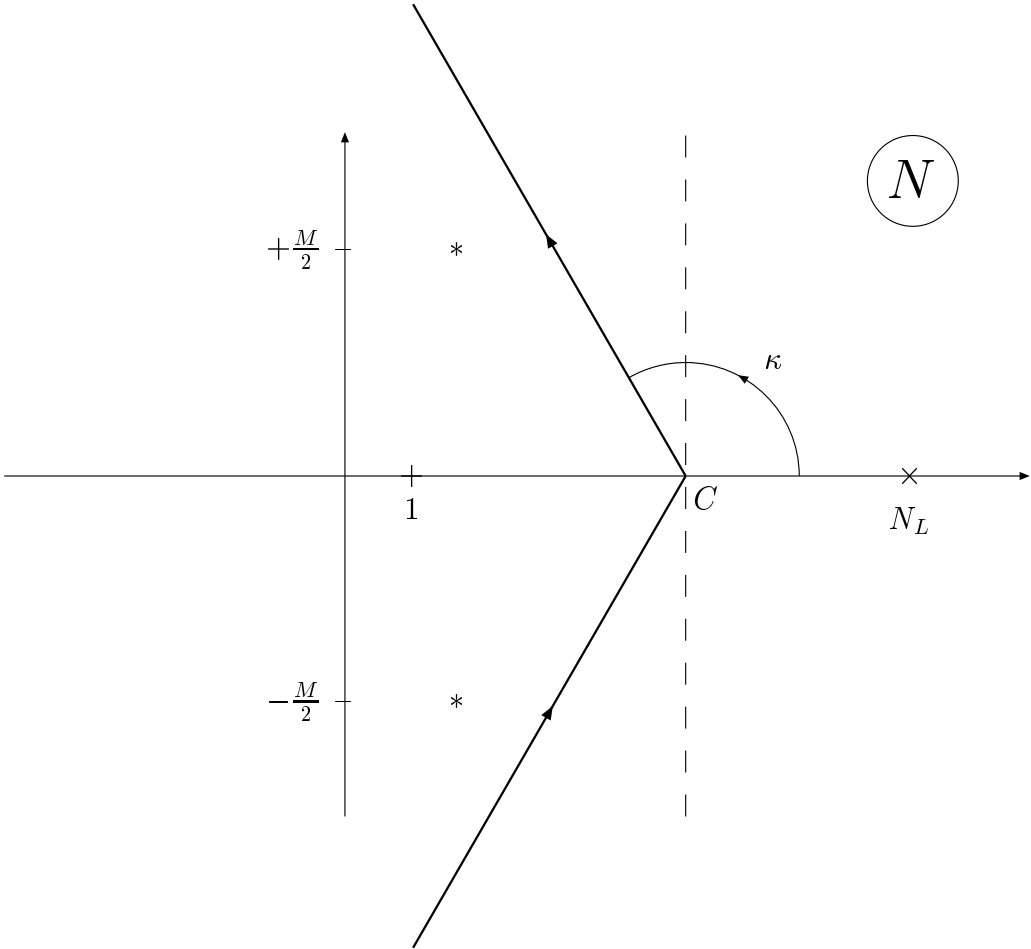,width=18.0cm}
\vspace*{-11cm}
\caption{Contour in Mellin-$N$ space for inverting the resummed cross
section. The asterisks denote the rightmost poles of the parton densities
which have acquired  an imaginary part through the Fourier transformation
in rapidity. $N_L$ is the position of the  
Landau pole; see text. \label{contour}}
\end{figure}

The final step is to insert the
result in Eq.~(\ref{etaexp}) back into (\ref{smn}), and to take the 
Mellin and Fourier inverse transforms of $\sigma (N,M)$, in order to arrive at 
the physical hadronic cross section:
\begin{eqnarray} \label{inverse} 
\frac{p_T^3 \d\sigma}{\d p_T \d\eta}
&=&\frac{1}{\sqrt{2 \pi}} \int_{-\infty}^{\infty} dM \, \e^{-i M \eta} \, 
\frac{1}{2 \pi i} \int_{C-i \infty}^{C+i \infty} dN \, (x_T^2)^{-N} \,
\sigma(N,M) \nonumber \\
&=&\frac{\alpha\as}{2 \pi} \int_{-\infty}^{\infty} dM \, \e^{-i M \eta} \, 
\frac{1}{2 \pi i} \int_{C-i \infty}^{C+i \infty} dN \, (x_T^2)^{-N} \,
\sum_{a,b} \tilde{\phi}_a^{N+1+\frac{i M}{2}}\tilde{\phi}_b^{N+1-\frac{i M}{2}}
\nonumber \\
&&\times\exp \left[ \ln \tilde{N} 
g_{ab}^{(1)} (\tilde{\lambda}) + g_{ab}^{(2)} \left(\tilde{\lambda},
\frac{Q^2}{\mu_R^2};\frac{Q^2}{\mu_F^2}\right)\right] \nonumber \\
&&\times\int_{-\infty}^{\infty} d\hat{\eta} \, \e^{i M \hat{\eta}} 
\left( \cosh \hat{\eta} \right)^{-2 N + g_{ab}^{(3)} (\tilde{\lambda})}
\left[ C_{ab}^{(0)} (\hat{\eta}) + \frac{\alpha_s}{\pi} 
 C_{ab}^{(1)} (\hat{\eta}) \right] \; ,
\end{eqnarray}
where for the second equality we have made use of Eqs.~(\ref{smn}) and
(\ref{etaexp}), and where the last integral is given explicitly
in Appendix~B, as discussed above.
Care has to be taken when choosing the contour in complex $N$ space.
First of all, the functions $g_{ab}^{(i)} (\lambda)$ have cut 
singularities starting at $\lambda=\frac{1}{2}$, that 
is, at $N=N_L=\exp (1/2b_0 \alpha_s (\mu_R^2))$.
These singularities result from the sensitivity of the original
resummed expression~(\ref{rescrosec}) to the Landau pole~\cite{CT}, 
and signal the onset of non-perturbative phenomena very close to threshold. 
With the ``minimal prescription''~\cite{CMNOV,MP} for the exponents
given in Eqs.~(\ref{g1fun})-(\ref{g3fun}) above, we choose  
the constant $C$ in~(\ref{inverse}) so that all singularities 
in the integrand are to the left of the integration contour, except for 
the Landau singularity at $N=N_L$, which lies to the far right.
The contour is then deformed~\cite{CMNOV,MP,ConSt} into the half-plane
with negative real part, which improves convergence while 
retaining the perturbative expansion. In this deformation, we need to 
avoid the moment-space singularities of the parton densities, which are 
displaced parallel to the imaginary axis by $\pm M/2$, as seen from 
Eq.~(\ref{smn}). Thus, the intersection at $C$ of the contour with
the real axis has to lie far enough to the right that the contour 
does not pass through or below the singularities of the parton densities. 
At the same time, one can make the bending angle $\kappa$ close to 
$90^{\circ}$ to prevent this from happening. Our parameterization of 
the contour is given as
\begin{eqnarray}
N &=& C + z \, \e^{\pm i \kappa} \; , \;\;\; (0\leq z  \leq \infty) \; ,
\nonumber \\
\kappa &=& \pi - \arctan \left( \frac{C-1+M/2}{C-1} \right) \; .
\end{eqnarray}
The situation is depicted in Fig.~\ref{contour}.

\section{Numerical results}
For our numerical calculations, we use the NLO parton densities 
of~\cite{grv}. The evolution code of~\cite{grv} is 
set up in Mellin-$N$ moment space and is therefore ideally 
suited for implementing the resummation formulas in the way we have
outlined above. In order to avoid double-counting of higher-order
corrections, we subtract from the resummed
expression~(\ref{rescrosec}) its perturbative expansion to
${\cal O}(\alpha_s)$, and subsequently add back in the full ``exact'' 
NLO prompt photon cross section~\cite{CMNOV}. For the latter, we use the 
calculation and the numerical code of~\cite{gv} and also 
include an NLO fragmentation component, computed using the results
of~\cite{aversa}, along with the NLO photon fragmentation functions
of~\cite{grvfrag}. Note that complete consistency would demand
to also take into account the effects of threshold resummation 
on the fragmentation component~\cite{CMNOV,LSV}. To mimic the expected 
effects, we multiply the NLO fragmentation component by the same
factor by which the non-fragmentation part is changed through threshold
resummation. Fortunately, the fragmentation contribution to the cross 
section is generally at the level of only $15-20 \%$ anyway~\cite{vv,CMNOV}, 
so the uncertainty residing here is not really crucial.

\begin{figure}[tbh]
\begin{center} 
\epsfig{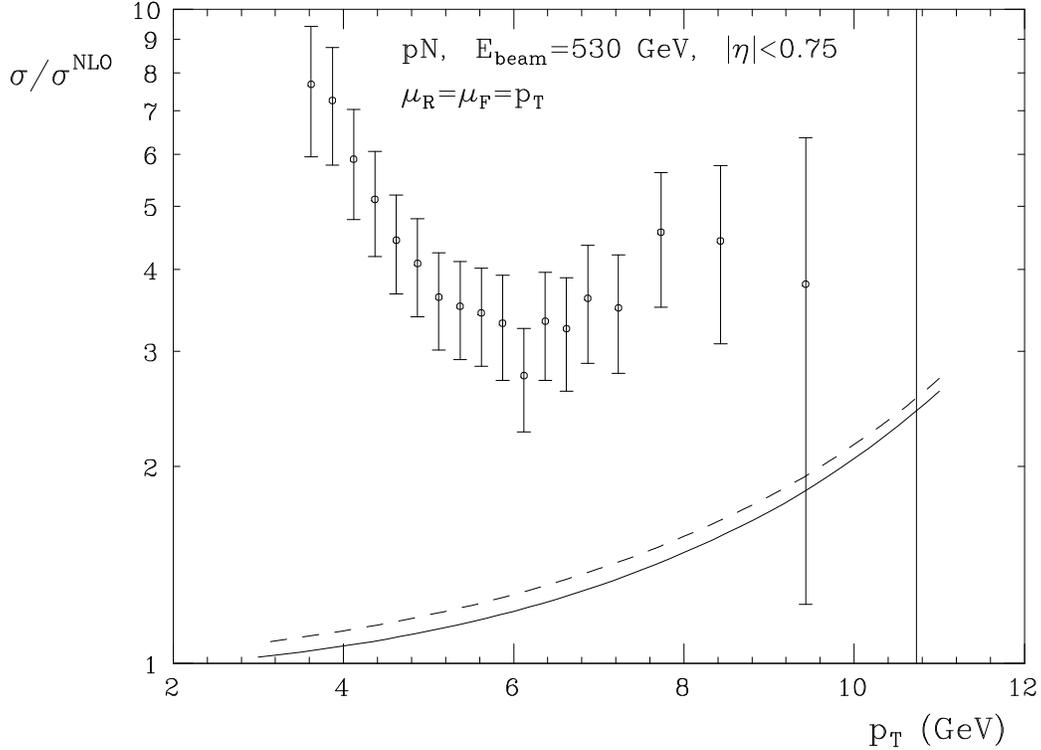}
\end{center}
\caption{Threshold-resummed prompt photon cross sections, normalized
to NLO, based on the formalisms of~\cite{CMN} (dashed) and~\cite{LOS}
(solid). The parton densities were taken from~\cite{grv}. The E706 prompt 
photon data~\cite{e706} are also shown in the same normalization.
\label{figpt}}
\end{figure}

As in~\cite{CMNOV,KO}, we will focus our calculations on the fixed-target
region, where the observed disagreement of data and NLO theory calculations
is largest, and which also is known~\cite{CMNOV,KO} to be generally more 
susceptible to threshold resummation effects than the collider regime, 
where the relevant values of $x_T$ are much smaller. Figure~\ref{figpt} 
compares our results for the $p_T$ dependence of the resummed 
prompt photon cross section with those of~\cite{CMNOV}. We use the same
parton distribution functions~\cite{grv} and rescaled fragmentation
contribution for both curves\footnote{The precise treatment of 
fragmentation and the choice of parton densities is different 
than in~\cite{CMNOV}. These choices have little numerical effect,
however.}. We show 
$E_{\gamma} d^3 \sigma/dp_{\gamma}^3$, calculated for p-N collisions at 
$E_{\rm beam}=530$ GeV, corresponding to the Fermilab E706 fixed-target 
experiment~\cite{e706}, whose data are also displayed. We have chosen the 
scales $\mu_F=\mu_R=p_T$ and integrated over the rapidity range $|\eta|\leq 
0.75$ accessed in the E706 experiment~\cite{e706}. For better visibility, we 
have normalized all cross sections by the NLO one. One notices first of all 
that threshold resummation induces large effects at high $p_T$, where it 
leads to a sizeable enhancement of the theory prediction~\cite{CMNOV,KO}, 
but is unimportant at the low-$p_T$ end. Figure~\ref{figpt} also shows that 
there are only small differences between the results given by the two 
resummation formalisms. These have two origins: first, they are related to 
the neglect of the rapidity dependence in the resummation exponent and in 
the higher-order terms of the $N$-independent coefficients $C_{ab}$ 
in~\cite{CMN}. As mentioned above, these differences 
would even be there, had we integrated over {\em all} kinematically 
allowed $\eta$, that is, over $|\eta| \leq \ln \left[ 
\left( 1+\sqrt{1-x_T^2} \, \right)/x_T \right]$.
Second, since rapidity dependence in resummation was not available 
in~\cite{CMNOV}, the means to compare to experimental data covering 
only a limited range of rapidity was to apply the following acceptance
correction to the resummed cross section:
\be \label{accept}
\sigma^{\rm (res)}(|\eta| \leq 0.75) 
\equiv \sigma^{\rm (res)}({\rm all}~ \eta) \;
\frac{\sigma^{\rm (NLO)}(|\eta| \leq 0.75)}
{\sigma^{\rm (NLO)}({\rm all}~ \eta)} \; .
\ee                                                        
That is, it was assumed that rescaling the $\eta$-integrated resummed cross 
section by an appropriate ratio of NLO cross sections correctly 
reproduces its rapidity dependence. This assumption of similarity of the 
resummed and NLO $\eta$-shapes is actually a fairly good approximation, as we 
will see below. We conclude from Fig.~\ref{figpt} that retaining the 
rapidity dependence in resummation induces very mild overall corrections 
for a cross section that is integrated over a substantial region including
$\eta=0$. 
As is evident from Figure~\ref{figpt}, threshold resummation effects,
albeit sizable in certain kinematic regions, are not sufficient to obtain 
a satisfactory description of the E706 data. This conclusion was first 
drawn in Refs.~\cite{CMNOV,KO}.

\begin{figure}[tbh]
\begin{center} 
\epsfig{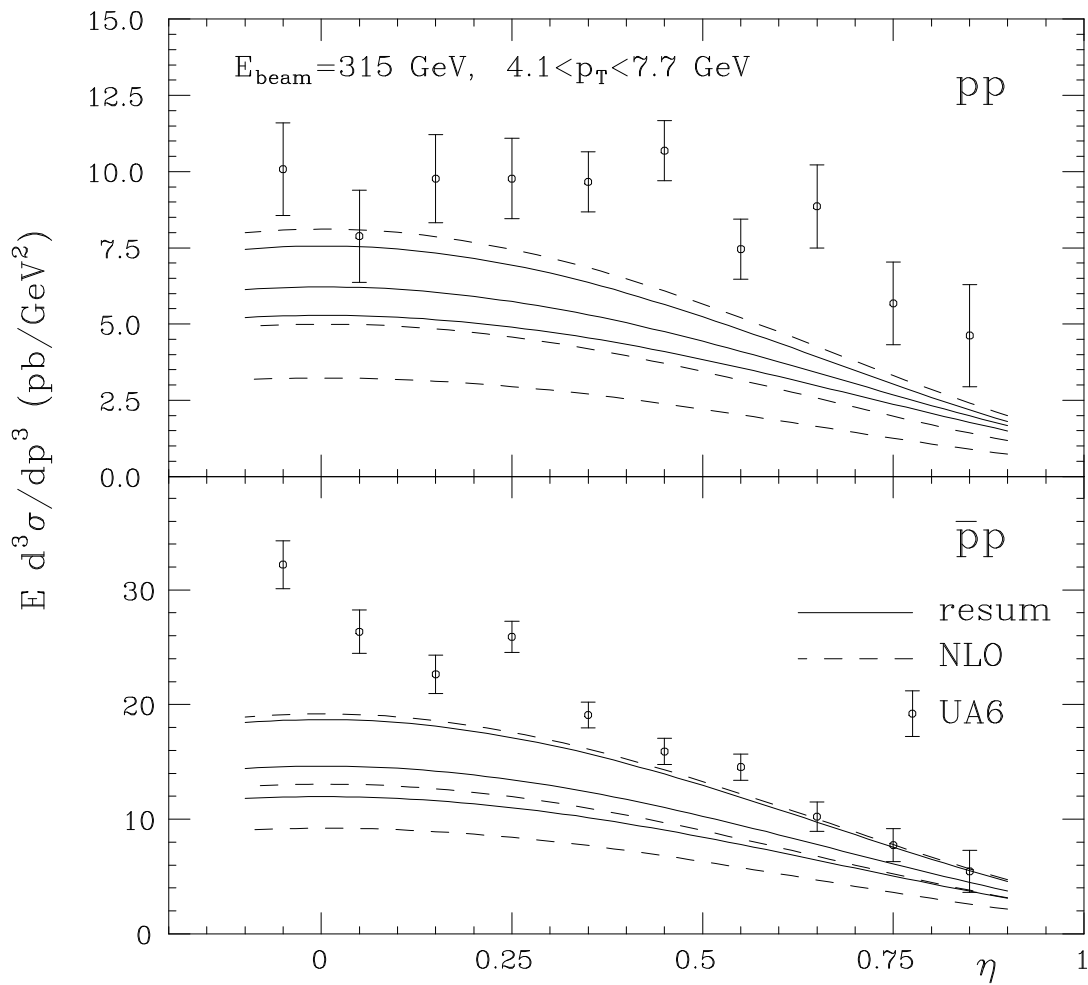}
\end{center}
\caption{Rapidity dependence of the prompt photon cross section 
in pp and $\bar{\rm p}$p collisions at $\sqrt{S}=24.3$ GeV. Dashed
lines are NLO, solid lines denote the cross section resummed to NLL
accuracy. For each case, the results have been calculated for
three choices of scales, $\mu_F = \mu_R = 2 p_T, p_T,p_T/2$, from 
lower to upper. The data are from UA6~\cite{ua6}.
\label{figpt2}}
\end{figure}

\begin{figure}[tbh]
\begin{center} 
\epsfig{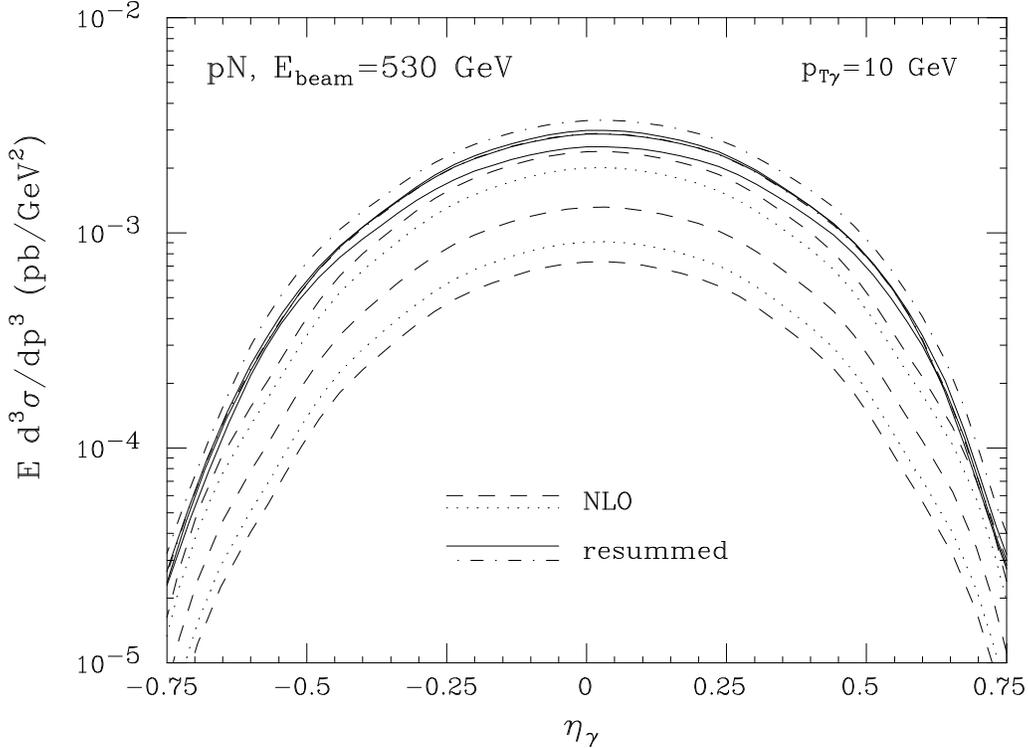}
\end{center}
\caption{Rapidity dependence of the prompt photon cross section 
in pN collisions at $\sqrt{S}=31.5$ GeV. Dashed and dotted lines are NLO, 
solid and dot-dashed lines denote the cross section resummed to NLL
accuracy. For each case, the results have been calculated for
five choices of scales, $\mu_F = \mu_R = 2 p_T, p_T,p_T/2$ (dashed and
solid), and $\mu_F = 2 p_T, p_T/2$ at fixed $\mu_R=p_T$ (dotted and
dot-dashed), from lower to upper.
\label{fig_530}}
\end{figure}

We now turn to the rapidity dependence of the threshold-resummed 
cross section, where using the full formalism of~\cite{LOS} presented above
is mandatory. The UA6 collaboration has presented data for the 
prompt photon cross section as a function of $\eta$, taken in pp and 
$\bar{\rm p}$p fixed-target scattering at $\sqrt{S}=24.3$ GeV~\cite{ua6}.
The transverse momentum of the photon was averaged over $4.1<p_T<7.7$ GeV.
In Fig.~\ref{figpt2} we compare our calculations to these data. In both 
plots, the dashed lines display the fixed-order NLO cross sections for three 
different choices of the renormalization and factorizations scales, 
$\mu_F = \mu_R = 2 p_T, p_T,p_T/2$, from lower to upper.
As can be seen, the variation of the NLO predictions induced by 
scale changes is very large, indicating the importance of yet higher-order 
effects. The solid lines in Fig.~\ref{figpt2} show the results after 
adding in the effects of NLL threshold resummation. The most striking 
feature is a strong reduction of scale dependence. This effect
was also found for the $\eta$-integrated cross section in~\cite{CMNOV};
our results demonstrate that it occurs over the whole $\eta$ range.
A closer analysis reveals that the main reduction in scale dependence is
associated with the factorization scale. The reasons for this were discussed
in~\cite{BCMN,SV}. Also, the scale dependence decreases even further if we
push our calculation to larger $x_T$. Both these observations are corroborated 
by Fig.~\ref{fig_530}, where we return to the conditions of the E706
experiment~\cite{e706}, looking at the rapidity dependence of the 
cross section at a fixed, high, photon transverse momentum $p_T=10$ GeV.
The dashed lines again show the NLO cross section for three choices
of scales, $\mu_F = \mu_R = 2 p_T, p_T,p_T/2$, and the solid lines 
display their resummed counterparts. Clearly, the reduction in scale
dependence after resummation is dramatic here. The dotted lines in
Fig.~\ref{fig_530} refer to the NLO cross section for fixed 
renormalization scale $\mu_R = p_T$, but varying the factorization
scale in the range used previously. The scale dependence of the NLO
cross section thus mainly is due to that on $\mu_F$, and the improvement
in $\mu_F$ dependence after resummation, shown by the dot-dashed lines,
is striking.

We note that resummation of the $\eta$-dependence of the 
cross section was also considered in~\cite{KO}. There the resummed 
expression was expanded to ${\cal O} (\alpha_s^2)$, thus testing the 
effects of the
NNLO terms generated by resummation. The numerical results of~\cite{KO} 
indeed also show some reduction in scale dependence which, however, is not 
as substantial as the one we find. This implies that resummation effects 
beyond NNLO are still important, and that the full resummed series should be 
taken into account. Our examples in Figs.~\ref{figpt2}, \ref{fig_530} 
also show that threshold resummation mainly affects the normalization of the 
cross section, and less so its shape in $\eta$ for moderate $\eta$.
Nevertheless, as we now show, large rapidity can, like high $p_T$, 
enhance resummation effects.

\begin{figure}[tbh]
\begin{center} 
\epsfig{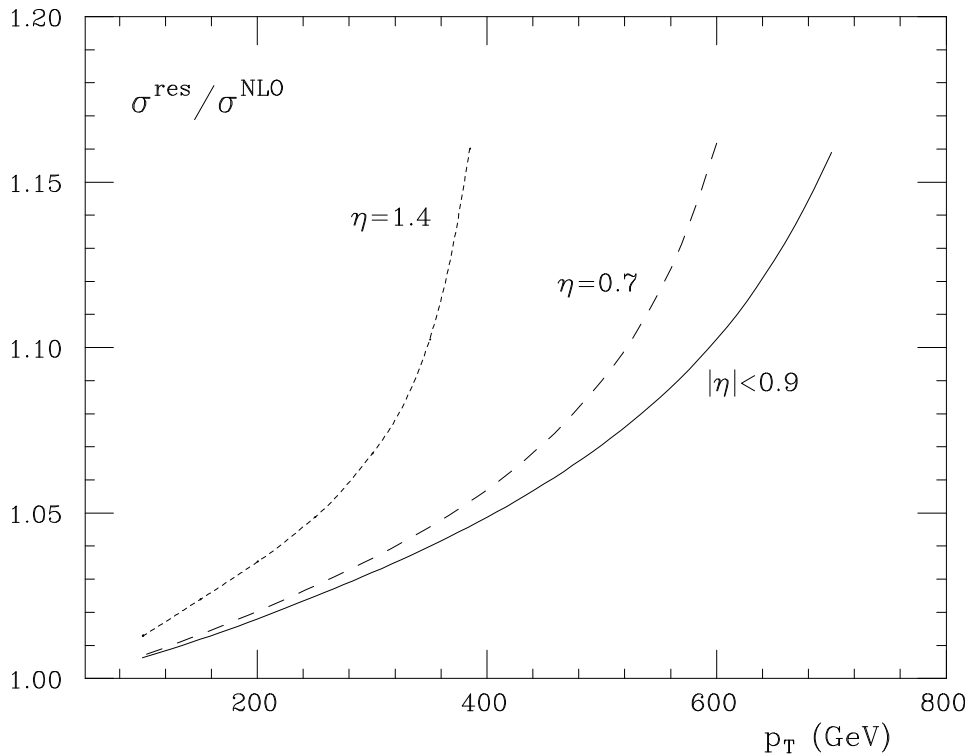}
\end{center}
\caption{Threshold-resummed prompt photon cross sections, normalized
to NLO, for p$\bar{\rm p}$ collisions at $\sqrt{S}=1800$ GeV.
\label{figcdf}}
\end{figure}

We have mentioned earlier that threshold resummation for the prompt photon
cross section is typically expected to be relevant in the fixed-target regime,
but less so for collider energies, where a given $x_T$ implies a 
much larger value of $p_T$ and hence a much smaller cross section. As a 
result, the highest-$p_T$ prompt photons seen at the Tevatron are at around
$p_T\approx 100$ GeV~\cite{cdf}, where $x_T\approx 0.1$ which is too 
small for threshold effects to be really important. However, we can ask
what the effects of threshold resummation would be, {\em were} we able
to go out in $p_T$ to, say, $p_T=500$ GeV. Keeping in mind that the 
Tevatron jet data~\cite{cdfjets} do reach out to $p_T$'s of that size,
this question becomes of some relevance, provided threshold 
resummation effects for the prompt photon cross section have any resemblance 
to those for the jet cross section. Fig.~\ref{figcdf} shows a  
calculation of threshold resummation effects on the cross section
for very-high-$p_T$ prompt photons, produced in p$\bar{\rm p}$ collisions 
at $\sqrt{S}=1800$ GeV. The results are normalized to the corresponding 
NLO cross section. The solid line shows the cross section averaged over
$|\eta|<0.9$, used for the actual CDF prompt photon data~\cite{cdf}, 
whereas the dashed and dotted lines present the result at a fixed near-central 
($\eta=0.7$) and forward ($\eta=1.4$) rapidity, respectively.
It is evident that at such high $p_T$ threshold resummation effects are 
sizable even at collider energy, and that, at a given $p_T$, they tend to 
become more important with increasing $|\eta|$, that is, toward
the limit of phase space where large momentum fractions in the parton 
densities are probed. At low $\eta$, effects remain modest even at 
large $p_T$, as recently observed directly in an NNLO expansion~\cite{KOjet} 
of the resummed jet cross section. Our finding could be indicative of 
significant effects also for jet production at the Tevatron, in particular 
in the off-central region. This result is also suggestive of 
recent data~\cite{d0new} which hints at an excess of high-$p_T$ jets that
increases with $\eta$. This issue will bear further investigation.

\section*{Acknowledgments}
We are grateful to S.~Catani and N.~Kidonakis for very valuable discussions. 
This work was supported in part by the National Science Foundation, grant
PHY9722101. G.S. acknowledges the hospitality of
Brookhaven National Laboratory.  
W.V. thanks RIKEN, Brookhaven National Laboratory and the U.S.
Department of Energy (contract number DE-AC02-98CH10886) for
providing the facilities essential for the completion of this work.

\section*{Appendix A}
In this appendix we give explicit expansions of the exponents
in Eq.~(\ref{rescrosec}) to NLL accuracy. These are analogous to 
the expressions given in Appendix~A of Ref.~\cite{CMN}.  
We expand each of the functions ${\cal G}_i(\as)\equiv
A_i(\as)$, $\bar B_i (\ldots,\as)$, 
$\gamma_i (\as)$, $\gamma_{ii} (\as)$, ${\rm Re}\Gamma_S^{(ab\to  \gamma c)}
(\hat{\eta},\as)$ as a power series in $\as/\pi$:
\begin{equation}
{\cal G}_i(\as) = \sum_{k=1}^{\infty} \left( \frac{\as}{\pi} \right)^k
{\cal G}_i^{(k)} (\as) \; .
\end{equation}
We then find
\begin{eqnarray}
\label{lndeltams}
&&E_i (2N_i,M_i)=\frac{1}{\alpha_s (\mu_R^2)}
{\cal H}_i^{(0)} (\lambda_i) + {\cal H}_i^{(1)} 
\left(\lambda_i,\frac{M_i^2}{\mu_R^2},
\frac{M_i^2}{\mu_F^2}\right) -\frac{1}{2\pi b_0} 
\bar{B}_i^{(1)} \left(\nu_i,\frac{M_i^2}{s} \right) \ln(1-2 \lambda_i)
\;  , \nonumber \\
&&E'_i (2N,s) =\frac{1}{\alpha_s (\mu_R^2)}
{\cal F}_i^{(0)} (\lambda) + {\cal F}_i^{(1)} 
\left(\lambda,\frac{s}{\mu_R^2}\right) -
\frac{1}{2\pi b_0} 
\bar{B}_i^{(1)} \left(\nu_i,1\right) \ln(1-2 \lambda)
\;  , \nonumber \\
&&\int_{\mu_R}^{\sqrt{s}/2N} {d\mu' \over \mu'} \, 
2 \, {\rm Re} \Gamma_S^{(ab\to  \gamma c)}
\left(\hat \eta,\alpha_s(\mu'^2)\right) =
\frac{\Gamma_S^{(ab\to  \gamma c)(1)}(\hat{\eta})}{\pi b_0} \;
\ln (1-2\lambda) 
\; ,
\end{eqnarray}
where $\lambda=\as (\mu_R^2) b_0 \ln N$, $\lambda_i=\as (\mu_R^2) b_0 
\ln N_i$, and
\begin{eqnarray}
{\cal H}_i^{(0)} (\lambda) &=& \frac{A_i^{(1)}}{2\pi b_0^2}
\Big[ 2 \lambda + (1 - 2 \lambda) \ln(1-2 \lambda) \Big]\, , 
\label{hsubadef0}\\
{\cal H}_i^{(1)} \left(\lambda,\frac{M_i^2}{\mu_R^2},
\frac{M_i^2}{\mu_F^2}\right)&=&
\frac{A_i^{(1)} b_1}{2\pi b_0^3} \left[ \frac{1}{2} \ln^2 (1-2 \lambda) +
2 \lambda + \ln(1-2 \lambda)  \right] -\frac{A_i^{(1)}(\GE+\ln 2)}{\pi b_0} 
\ln(1-2\lambda) \nonumber \\
&&+ \; \frac{1}{2\pi b_0} \left( - \frac{A_i^{(2)}}{\pi b_0} +
A_i^{(1)} \ln \left( \frac{M_i^2}{\mu_R^2} \right) \right)
\Big[ 2 \lambda + \ln(1-2 \lambda) \Big] -
\frac{A_i^{(1)}}{\pi b_0} \lambda \ln \left( \frac{M_i^2}{\mu_F^2}
\right) \; ,\nonumber \\
{\cal F}_i^{(0)} (\lambda) &=& 2 {\cal H}_i^{(0)} (\lambda/2) -  
{\cal H}_i^{(0)} (\lambda)\; , \nonumber \\
{\cal F}_i^{(1)} (\lambda) &=& 2 {\cal H}_i^{(1)} (\lambda/2) -  
{\cal H}_i^{(1)} (\lambda) 
+\frac{1}{\pi b_0} \left[ A_i^{(1)} (\GE + \ln 2 ) -
\gamma_i^{(1)} \right] \ln(1-\lambda) \; .
\label{hsubadef} 
\end{eqnarray}
To NLL accuracy, we further expand
\begin{equation}
\ln (1-2 \lambda_i) \approx \ln(1-2 \lambda) - \frac{2\; 
\as(\mu_R^2)\; b_0}{1-2 \lambda}
\ln (N_i/N) \; ,
\end{equation}
where we recall that $N_a/N=-u/s$, $N_b/N=-t/s$. From the appropriate 
sums of these
expressions, the $g^{(i)}_{ab}$'s of Eqs.~(\ref{g1fun})-(\ref{g3fun1})
are identified through the linear dependence on $\ln N$ and 
$\ln \left( \cosh \hat{\eta}\right)$. 

\newpage
\section*{Appendix B}
In this appendix, we exhibit the $\hat{\eta}$-integrals of 
Eq.~(\ref{etaexp}) for the coefficients $C_{ab}^{(0,1)}(\hat{\eta})$ of 
Eqs.~(\ref{clo}), (\ref{c1qqb}), (\ref{c1qg}).
After substitution $x\equiv \exp (2\hat{\eta})$, 
one is led to integrals of the type 
\beq
P_{c,d}(a,b) \equiv 2^{-2-b} 
\int_0^{\infty} dx \, x^a \, (1+x)^b \, \ln^c x \, \ln^d (1+x) \;\; 
\;\;\; (c,d \in \{ 0,1,2 \}) \; ,
\eeq
which can be expressed in terms of the $\Gamma$ function and its
derivatives:
\begin{eqnarray} \label{psifn}
P_{0,0} (a,b) &=& \frac{1}{2} \;B \left( \frac{1}{2} ,-\frac{b}{2} \right) 
\frac{\Gamma (a+1)\Gamma (-a-b-1)}{\Gamma^2 (-b/2)} \nonumber
\; , \\
P_{1,0} (a,b) &=&P_{0,0} (a,b)\; \left[ \psi (a+1) - \psi (-a-b-1) \right]
\; ,\nonumber \\
P_{0,1} (a,b) &=&P_{0,0} (a,b)\; \left[ \psi (-b) - \psi (-a-b-1) \right]
\; ,\nonumber \\
P_{2,0} (a,b) &=&P_{0,0} (a,b)\; \left[ \psi' (a+1) + \psi' (-a-b-1) \right]
+ \left[ P_{1,0} (a,b) \right]^2 /P_{0,0} (a,b)
\; ,\nonumber \\
P_{0,2} (a,b) &=&P_{0,0} (a,b)\; \left[ \psi' (-a-b-1) -\psi' (-b) \right]
+\left[ P_{0,1} (a,b) \right]^2 /P_{0,0} (a,b) \; ,\nonumber \\
P_{1,1} (a,b) &=&P_{0,0} (a,b)\; \psi' (-a-b-1) +
P_{1,0} (a,b) P_{0,1} (a,b) /P_{0,0} (a,b)
\; ,
\end{eqnarray}
where $\psi(z) \equiv d\ln\Gamma (z) /dz$ and $\psi'(z)\equiv d\psi(z)/dz$. 
The numerical treatment of these functions for complex arguments 
was discussed in~\cite{grvold}. For convenience, we also abbreviate 
\begin{eqnarray}
R_{i,j} (a,b) &\equiv& P_{i,j} (a,b) + P_{i,j} (a+2,b) \; , \nonumber\\
S_{i,j} (a,b) &\equiv& 2 P_{i,j} (a,b) + 2 P_{i,j} (a+1,b) + 
P_{i,j} (a+2,b) \; .
\end{eqnarray}
We then have for $q\bar{q}$ scattering:
\begin{eqnarray} \label{cnqqb} 
&&\int_{-\infty}^{\infty} d\hat{\eta} \, \e^{i M \hat{\eta}} 
\left( \cosh \hat{\eta} \right)^{-2 N + g_{q\bar{q}}^{(3)} (\tilde{\lambda})}
\left[ C_{q\bar{q}}^{(0)} (\hat{\eta}) + \frac{\alpha_s}{\pi} 
 C_{q\bar{q}}^{(1)} (\hat{\eta}) \right]  \nonumber \\
&=&\frac{C_F \pi}{C_A} e_q^2 \left\{ R_{0,0} (a,b) 
\Bigg( 1 + \frac{\alpha_s}{\pi} \Bigg[ (2 C_F-C_A/2) 
\left(\gamma_E^2 + 2 \zeta(2) \right)+ \pi b_0 (\gamma_E + 2 \ln 2)
\right. \nonumber \\
&&\hspace*{4.6cm}
+(6 C_F-C_A) \gamma_E \ln 2- \frac{3}{2} C_F \ln 2
+ (4 C_F-C_A/2) \ln^2 2  \nonumber \\
&&\left. \left. \hspace*{4.55cm}+ K/2 -K_q - C_F (2 \gamma_E+2 \ln 2-3/2) \ln 
\frac{Q^2}{\mu_F^2} - \pi b_0  \ln \frac{Q^2}{\mu_R^2} \right] \right)
\nonumber \\
&&+\frac{\alpha_s}{\pi}\Bigg\{ -\frac{3}{2} C_F R_{0,1} (a,b) 
- \frac{1}{4} \Big( R_{2,0} (a,b) + 3 R_{0,2}\Big)
\left( 3 C_A - 10 C_F \right) + \frac{1}{2} R_{1,1} (a,b) 
\left( 5 C_A - 16 C_F \right)\nonumber \\
&&+\Big(R_{1,0} (a,b) - 2 R_{0,1} (a,b) \Big)
\left( -\frac{3}{2} C_F +
(4 C_F-C_A) \gamma_E + \pi b_0 - C_F \ln \frac{Q^2}{\mu_F^2} 
+(5 C_F-C_A) \ln 2 \right)\nonumber \\
&& + \frac{3}{2} C_F P_{1,0} (a,b) 
+ \frac{1}{4} (C_A - 2 C_F ) \Big( - 2 P_{1,0} (a+1,b) 
+ 4  P_{0,1} (a+1,b) + P_{2,0} (a,b)
\nonumber \\
&& - 2 P_{2,0} (a+1,b) - 4 P_{0,2} (a+1,b) -
2 P_{1,1} (a,b) + 4 P_{1,1} (a+1,b) \Big) \Bigg\}\Bigg\} \; ,
\end{eqnarray}
where
\begin{eqnarray}
K_q &=& \left( \frac{7}{2} - \zeta(2) \right) C_F \; ,  \nonumber \\
a&=& N-g_{q{\bar q}}^{(3)}(\tilde{\lambda})/2 +i M/2 \; ,  \nonumber \\
b&=& -2 N +g_{q{\bar q}}^{(3)}(\tilde{\lambda})-4 \; .
\end{eqnarray}
The coefficient $K$ has already been defined in Eq.~(\ref{anuf}). 
For $qg$ scattering, we find:
\begin{eqnarray}\label{cnqg}
&&\int_{-\infty}^{\infty} d\hat{\eta} \, \e^{i M \hat{\eta}} 
\left( \cosh \hat{\eta} \right)^{-2 N + g_{qg}^{(3)} (\tilde{\lambda})}
\left[ C_{qg}^{(0)} (\hat{\eta}) + \frac{\alpha_s}{\pi} 
 C_{qg}^{(1)} (\hat{\eta}) \right] \\
&=&\frac{\pi}{4C_A} e_q^2 \left\{ S_{0,0} (a,b)
\Bigg( 1 + \frac{\alpha_s}{\pi} \Bigg[ (C_A+C_F/2) \gamma_E^2 
-\frac{1}{2} K_q + (C_A+2C_F) \zeta(2)+ (2 C_A+3 C_F/2) \ln^2 2
\right. \nonumber \\
&&\left. \left.
+(3 C_A+2 C_F) \gamma_E \ln 2 + \pi b_0 \ln \frac{\mu_R^2}{\mu_F^2}
+\frac{3}{4} C_F \left (\gamma_E + \ln \frac{Q^2}{\mu_F^2} \right)
-(C_A+C_F) (\gamma_E + \ln 2) \ln \frac{Q^2}{\mu_F^2}\right] \right)
\nonumber \\
&&+\frac{\alpha_s}{\pi}\Bigg\{ S_{1,0} (a,b) 
\left(-\frac{3}{4} C_F + (2 C_A+C_F) \gamma_E + 2 ( C_A + C_F) \ln 2
- C_F \ln \frac{Q^2}{\mu_F^2} \right) \nonumber \\
&&-S_{0,1}\left(2  (2C_A+C_F) \gamma_E  +(5 C_A +3C_F) \ln 2
- (C_A+C_F) \ln \frac{Q^2}{\mu_F^2} \right) +
\frac{1}{2} (C_A+C_F) P_{0,1} (a,b) \nonumber \\
&&- \frac{1}{4} (C_A - 8 C_F) 
S_{2,0} (a,b) + \frac{1}{4} (11C_A +10 C_F) 
S_{0,2} (a,b) - \frac{1}{2} (3C_A +8 C_F) 
S_{1,1} (a,b) \nonumber \\
&&+\frac{1}{4} (C_A-2C_F) \Big( (P_{0,0} (a,b) + 2 P_{0,0} (a+1,b) ) 
\pi^2 - 4 P_{1,0} (a,b) -4 P_{1,0} (a+1,b) +2 P_{0,1} (a+1,b)\nonumber \\
&&+  4 P_{2,0} (a,b) +4 P_{2,0} (a+1,b)
+ 3 P_{0,2} (a,b) 
+ 2 P_{0,2} (a+1,b) - 6 P_{1,1} (a,b) - 4 P_{1,1} (a+1,b) \Big) 
\Bigg\}\Bigg\} \; , \nonumber
\end{eqnarray}
where in this case
\begin{eqnarray}
a&=& N-g_{qg}^{(3)}(\tilde{\lambda})/2 +1+i M/2\; ,  \nonumber \\
b&=& -2 N +g_{qg}^{(3)}(\tilde{\lambda})-5 \; .
\end{eqnarray}
It is evident from Eq.~(\ref{psifn}) that when performing 
the $\hat{\eta}$ integrations one inevitably introduces non-leading
contributions in the moment variable $N$. This is already true
when keeping only the Born approximations $C_{ab}^{(0)}(\hat{\eta})$
for the hard scattering, and is not in conflict with the 
approximations made in threshold resummation, where the resummed 
exponents are always convoluted with smooth functions, like the parton
densities, that abundantly contain terms subleading in $N$.
We note that for large $N$ the one-loop parts of 
Eqs.~(\ref{cnqqb}), (\ref{cnqg}) converge to the respective coefficients
given in Eqs. (58), (59) of~\cite{CMN} for the case of the 
$\eta$-integrated cross section, as they must.

\end{document}